\documentclass[submission, Phys]{SciPost}

\usepackage[english]{babel}
\usepackage[utf8]{inputenc}
\usepackage[T1]{fontenc}
\usepackage{pdfpages}
\usepackage{comment}

\usepackage{amsmath, amsthm, amssymb}
\usepackage{mathrsfs}
\usepackage{siunitx}
\usepackage{float}
\usepackage{graphicx}
\usepackage[colorinlistoftodos]{todonotes}
\usepackage{tikz}
\usepackage{braket}
\usepackage{listings}

\usepackage{caption}
\usepackage{subcaption}

\usepackage{dsfont}

\allowdisplaybreaks

\begin{document}

\begin{center}{\Large \textbf{
	Projective cluster-additive transformation for\\ quantum lattice models
}}\end{center}

\begin{center}
Max Hörmann\textsuperscript{1$\star$} and
Kai Phillip Schmidt\textsuperscript{1$\diamond$}
\end{center}

\begin{center}
{\bf 1} {Department of Physics, Staudtstra{\ss}e 7, Friedrich-Alexander-Universit\"at Erlangen-N\"urnberg (FAU), Germany}

${}^\star$ {\small \sf max.hoermann@fau.de}\\
${}^{\diamond}$	{\small \sf kai.phillip.schmidt@fau.de}
\end{center}

\begin{center}
\today
\end{center}


\section*{Abstract}
{
We construct a projection-based cluster-additive transformation that block-diagonalizes wide classes of lattice Hamiltonians $\mathcal{H}=\mathcal{H}_0 +V$. Its cluster additivity is an essential ingredient to set up perturbative or non-perturbative linked-cluster expansions for degenerate excitation subspaces of $\mathcal{H}_0$. Our transformation generalizes the minimal transformation known amongst others under the names Takahashi's transformation, Schrieffer-Wolff transformation, des Cloiseaux effective Hamiltonian, canonical van Vleck effective Hamiltonian or two-block orthogonalization method. The effective cluster-additive Hamiltonian and the transformation for a given subspace of $\mathcal{H}$, that is adiabatically connected to the eigenspace of $\mathcal{H}_0$ with eigenvalue $e_0^n$, solely depends on the eigenspaces of $\mathcal{H}$ connected to $e_0^m$ with \mbox{$e_0^m\leq e_0^n$}. In contrast, other cluster-additive transformations like the multi-block orthognalization method or perturbative continuous unitary transformations need a larger basis. This can be exploited to implement the transformation efficiently both perturbatively and non-perturbatively. As a benchmark, we perform perturbative and non-perturbative linked-cluster expansions in the low-field ordered phase of the transverse-field Ising model on the square lattice for single spin-flips and two spin-flip bound-states.
}

\vspace{10pt}
\noindent\rule{\textwidth}{1pt}
\tableofcontents\thispagestyle{fancy}
\noindent\rule{\textwidth}{1pt}
\vspace{10pt}

\section{Introduction}
In order to solve the time-independent Schrödinger equation for a Hamiltonian on a lattice
\begin{equation}
	\mathcal{H} = \mathcal{H}_0 + \lambda V
	\label{eq::define Hamiltonian}
\end{equation}
one needs to find the eigenvalues and eigenfunctions of $\mathcal{H}$. In many situations the part $\mathcal{H}_0$ is easy to solve and can be written in diagonal form, while 
\begin{equation}
	[\mathcal{H}_0,V] \neq 0
	\label{eq::H non-commuting parts}
\end{equation}
makes solving $\mathcal{H}$ a difficult problem. Since in the generic case one can not find solutions for $\mathcal{H}$, one has to resort to approximations. One of the oldest is perturbation theory. While the first two orders of perturbation theory normally can be easily calculated by hand, high orders are only accessible with computer aid and several methods for their computation exist. Albeit many other numerical techniques exist nowadays, high-order series expansions are used as a competitive technique to tackle quantum many-body problems \cite{Knetter2000,Trebst2000,Zheng2006}. Examples range from the calculation of low- and high-field expansions for transverse-field Ising models \cite{He1990,Oitmaa1991}, the analysis of phase transitions in triangular-lattice bilayer Heisenberg models \cite{Singh1998a} and spectral densities of two-particle excitations in dimerized Heisenberg quantum spin systems \cite{Trebst2000,Zheng2001,Knetter2003b} to the study of critical and Griffiths-McCoy singularities in quantum Ising spin-glasses \cite{Singh2017} or the derivation of spectral densities for Heisenberg quantum magnets with quenched disorder \cite{Hormann2018,Hormann2020}, or to the analysis of quantum phase diagrams of long-range transverse-field Ising models \cite{Fey2016} and the application to quantum phases with intrinsic topological order \cite{Muehlhauser2020,Wiedmann2020,Muehlhauser2022b}. Also questions such as the exploration of possible ground states in the kagome Heisenberg model \cite{Singh2007} can be tackled with perturbation theory.\\
The necessity of a perturbative starting point is not only a drawback but also helps in giving a clear picture of the physical problem at hand. To analyse quantum phase diagrams one usually investigates how the phase present at $\lambda=0$ breaks down by applying extrapolation techniques to high-order series expansions of relevant energies or observables. The accuracy of those increases with higher orders of perturbation available. This shows that the efficiency of the method used to derive the perturbative expansion is crucial.\\
A common approach to calculate quantities perturbatively on a lattice is to do a graph decomposition. Instead of a large cluster the calculations are performed on many small ones, which decreases memory requirements and is easily parallelized. The calculated values of a quantity $M$ on the subgraphs of the lattice are then multiplied with embedding factors to obtain the value of $M$ up to a given order on the whole lattice making use of the inclusion-exclusion principle. If for two disconnected parts $A$ and $B$  of the lattice the operator $M(A\cup B)$ is the direct sum
\begin{equation}
	M(A\cup B) = M(A) \oplus M(B),
	\label{eq:: additivity quantitiy M}
\end{equation}
the graph expansion can be restricted to connected subgraphs of the lattice. An operator $M$ that fulfils property \eqref{eq:: additivity quantitiy M} is called additive. However, not every transformation yields an effective Hamiltonian that allows a decomposition of the form \eqref{eq:: additivity quantitiy M}. In particular, there is an efficient block-diagonalisation method, that only makes use of the projectors of eigenspaces of $\mathcal{H}_0$ and $\mathcal{H}$ and is known under different names, for example as Takahashi's transformation, Schrieffer-Wolff transformation, des Cloiseaux effective Hamiltonian, canonical van Vleck effective Hamiltonian or two-block orthogonalization method \cite{DesCloizeaux1960, Hasegawa1977,Shavitt1980,Cederbaum1989,Zheng2006}, which in general does not allow to perform calculations on linked subgraphs of the lattice only. This is unfortunate since this transformation can be efficiently calculated using matrix-vector multiplications only \cite{Zheng2006}.\\
Non-perturbative linked-cluster expansions (NLCEs) follow the same principles as perturbative expansions but use non-perturbative cluster results, which are in many cases just the exact results of the finite cluster. They were first introduced in \cite{Hamer1983} and were often used for thermodynamic quantities \cite{Rigol2007a} or ground-state expectation values \cite{Rigol2006}. For non-perturbative expansions it is even more important that the expansion can be performed on linked clusters only. Otherwise finding a hierarchy to truncate the expansion is difficult. For excited states linked-cluster expansions were performed with flow-equations in an approach called graph-based continuous unitary transformations (gCUT) \cite{Yang2011}. Another expansion, but only relying on the eigenvectors and energies of the block of interest, is the contractor renormalization group method (CORE) \cite{Morningstar1996}. In contrast to gCUT, it does not fulfil the linked-cluster property in general. However, a great advantage is its efficiency only relying on the low-energy eigenstates that can be calculated with numerical routines such as the Lanczos algorithm. The CORE method is similar to the projective transformation mentioned above but is not as efficient in the perturbative regime in the sense that it needs more clusters to converge to a given order of perturbation.\\
In this paper we will introduce an optimal transformation: It shares the efficiency of the projective method, can also be applied non-perturbatively using the exact lowest eigenvectors and energies, but also allows for a cluster expansions with linked clusters only. We do this by extending the projective transformation for an eigenspace adiabatically connected to $e_0^n$, where $e_0^n$ denotes the energy of the degenerate subspaces of $\mathcal{H}_0$, to incorporate eigenstates adiabatically connected to blocks $m$ with $e_0^m<e_0^n$ and not only those of $e_0^n$. Before describing the important changes to the transformation we review other approaches to construct a genuine linked-cluster transformation and inform about different equivalent formulations of the classical projective transformation in Sec.~\ref{section::Block-diagonalisation of Hamiltonians}. Then we exemplify the roots of the linked-cluster violation of the projective transformation with a simple toy model. In Sec.~\ref{section::Projective transformation fulfilling the linked-cluster property} we show how these problems can be cured for multi-particle excitations in general and also give a general form of the transformation in terms of projection operators. As an application, in Sec.~\ref{section::Application: low-field expansion for transverse-field Ising model on square lattice} we apply the method to the low-field expansion of the TFIM on the square lattice, both perturbatively and non-perturbatively. We conclude our work in Sec.~\ref{section::Conclusions}.
\section{Block-diagonalisation methods}
\label{section::Block-diagonalisation of Hamiltonians}
In this section, we first define what block-diagonal form we want to achieve with block-diagonalisation methods and fix basic notation. Then we review existing cluster-additive block-diagonalisation methods and the projective minimal transformation.
\subsection{Block-diagonalised form and cluster-additivity}
The Hilbert space $\mathscr{H}$ of finite dimension $N$ can be written as the direct sum of the eigenspaces $\mathscr{H}_0^n$ of the operator $\mathcal{H}_0$:
\begin{equation}
	\mathscr{H} = \bigoplus_{n=0}^N \mathscr{H}_0^n
	\label{eq::Hilbert space direct sum of H_0}
\end{equation}
As $\mathcal{H}_0$ is assumed to have block diagonal form we have
\begin{equation}
	\mathcal{H}_0 = \bigoplus_{n=0}^N \mathcal{H}_0^n,
	\label{eq::H_0 operator as direct sum}
\end{equation}
where the ordering of eigenvalues of the eigenspaces is $e_0^m \leq e_0^n$ for $ m\leq n$. 
In more explicit form the parts $\mathcal{H}_0^n$ fulfil
\begin{equation}
	\mathcal{H}_0 \,v= \left(\bigoplus_{n=0}^N \mathcal{H}_0^n\right) v = \left(\bigoplus_{n=0}^N \mathcal{H}_0^n \,v_{0,n}\right) 
	\label{eq::H_0^n action on states}
\end{equation}
for $v = \sum_{n=0}^{N} v_{0,n}$ and $v_{0,n}\in \mathscr{H}_0^n$.
For a block-diagonalising unitary transformation $T$ and the corresponding effective Hamiltonian \mbox{$\mathcal{H}_{\mathrm{eff}} = T^\dagger \mathcal{H} T^{\phantom\dagger}$}, unitarity implies
\begin{equation}
	\mathscr{H} = \bigoplus_{n=0}^N \mathscr{H}_{\mathrm{eff}}^n = \bigoplus_{n=0}^N T \mathscr{H}_0^n
	\label{eq::Hilbert space direct sum of H_eff}
\end{equation}
as well as $\mathcal{H}_{\mathrm{eff}}$ to be block-diagonal so that it can be written as
\begin{equation}
	\mathcal{H}_{\mathrm{eff}} = \bigoplus_{n=0}^N \mathcal{H}_{\mathrm{eff}}^n,
	\label{eq::operator H_eff as direct sum}
\end{equation}
i.e.
\begin{equation}
	\mathcal{H}_{\mathrm{eff}} \,v= \left(\bigoplus_{n=0}^N \mathcal{H}_{\mathrm{eff}}^n\right) v = \left(\bigoplus_{n=0}^N \mathcal{H}_{\mathrm{eff}}^n \,v_n\right) 
	\label{eq::definition direct sum of H_eff}
\end{equation}
for $v = \sum_{n=0}^{N} v_n$ and $v_n\in \mathscr{H}_{\mathrm{eff}}^n$.\\

After having defined the block-diagonalised form of the effective Hamiltonian \eqref{eq::operator H_eff as direct sum} resulting from a unitary transformation $T$, 
we next introduce the concept of cluster-additivity for such transformations. Historically, first linked-cluster expansions for perturbative ground-state energy calculations were performed in 1955 \cite{Brueckner1955} and applied to calculate zero-temperature ground state properties in high orders later in the 1980s using Nickel's cluster expansion method from unpublished work \cite{Marland1981,Hamer1983}. The transformation used to calculate ground-state properties is not important since the ground-state additivity 
\begin{equation}
	e^0(A\cup B) = e^0(A) + e^0(B) 
	\label{eq::cluster additivity ground-state energy}
\end{equation}
is always fulfilled for disconnected clusters $A$ and $B$ assuming a non-degenerate ground-state subspace. With Nickel's cluster expansion method even excitation gaps could be calculated \cite{He1990} by grouping terms in orders of the number of sites of the lattice, although a restriction to linked clusters was not sufficient for that. Still, these calculations were more efficient than calculations on linked-clusters using a cluster-additive transformation \cite{Dusuel2010} due to the higher efficiency of the method. The proper formalism to derive the right cluster-additive part of the effective one-particle Hamiltonian was written down in 1996 by Gelfand \cite{Gelfand1996}. A more extensive review can be found in \cite{Gelfand2000}. The decisive point was to not do a linked-cluster expansion for the effective Hamiltonian in the one-particle space $\mathcal{H}_{\mathrm{eff}}^1$ but to the effective Hamiltonian minus the ground-state energy:
\begin{equation}
	\bar{\mathcal{H}}_{\mathrm{eff}}^1\vert_1 (A\cup B) \equiv \mathcal{H}_{\mathrm{eff}}^1(A\cup B)-e^0(A\cup B) = \bar{\mathcal{H}}_{\mathrm{eff}}^1\vert_1 (A) \oplus \bar{\mathcal{H}}_{\mathrm{eff}}^1\vert_1 (B)
	\label{eq::cluster additivity one-particle effective Hamiltonian}
\end{equation}
This was generalized to a proper cluster expansion for two particles around 2000 \cite{Trebst2000,Knetter2000a,Zheng2001} and was further generalized to multi-particle excitations in 2003 \cite{Knetter2003a}. They introduced the notion of cluster additivity: An effective cluster additive Hamiltonian takes the form 
\begin{equation}
	\mathcal{H}_{\mathrm{eff}}(A\cup B) = \mathcal{H}_{\mathrm{eff}}(A) \otimes \mathds{1}_B + \mathds{1}_A \otimes \mathcal{H}_{\mathrm{eff}}(B)
	\label{eq::cluster additivity effective Hamiltonian}
\end{equation}
on disconnected parts $A$ and $B$ of the lattice. We stress that this form is different to the direct sum in Eq.~\eqref{eq:: additivity quantitiy M}. However, if the effective Hamiltonian takes the cluster-additive form of Eq.~\eqref{eq::cluster additivity effective Hamiltonian}, it can be decomposed into additive parts and a linked-cluster expansion can be performed. These additive parts, denoted by $\bar{\mathcal{H}}_{\mathrm{eff}}^n$, are inductively defined by
\begin{equation}
	\begin{aligned}
		\mathcal{H}_{\mathrm{eff}}^0&=\bar{\mathcal{H}}_{\mathrm{eff}}^0\\
		\mathcal{H}_{\mathrm{eff}}^1&=\bar{\mathcal{H}}_{\mathrm{eff}}^0\vert_1 + \bar{\mathcal{H}}_{\mathrm{eff}}^1\vert_1\\
		\vdots\\
		\mathcal{H}_{\mathrm{eff}}^N&=\sum_{n=0}^{N}\bar{\mathcal{H}}_{\mathrm{eff}}^n\vert_N.
	\end{aligned}
	\label{eq::proper reduced bar_H_eff^n}
\end{equation}
The first two equations are precisely what was described by Gelfand \cite{Gelfand1996}. To understand the action of $\bar{\mathcal{H}}_{\mathrm{eff}}^m\vert_n$ on a state one has to expand the state in the position basis. Then, for each position basis state, one finds all product state decompositions into two position basis states. $\bar{\mathcal{H}}_{\mathrm{eff}}^m\vert_n$ then acts with an identity on the one part of the product state having energy $e_0^n-e_0^m$ in $\mathcal{H}_0$, and with $\bar{\mathcal{H}}_{\mathrm{eff}}^{m}\vert_m$ on the other part.

\subsection{Cluster-additive block diagonalisation methods}
\label{subsection::Cluster-additive block diagonalisation methods}

The subtractions of Eq.~\eqref{eq::proper reduced bar_H_eff^n} are necessary to perform linked-cluster expansions but not sufficient. For degenerate subspaces of $\mathcal{H}_0$ the transformation used is not uniquely determined and the cluster-additivity property of \eqref{eq::cluster additivity effective Hamiltonian} is not necessarily given. There are two prominent approaches to construct cluster-additive effective Hamiltonians. Both make use of the linking structure of the commutator.\\ The first one are continuous unitary transformations (CUTs), which are defined by the flow equations
\begin{equation}
	[\eta,\mathcal{H}] = \partial_l \mathcal{H}
	\label{eq::flow equation}
\end{equation}
with $\eta(l)$ the anti-Hermitian generator of the transformation. In physics they were introduced 1993 by Wegner \cite{Wegner1994} and Glazek and Wilson \cite{Gazek1993} with the double-bracket flow, which was known in mathematics already in 1988 \cite{Brockett1988}. To use flow equations to study eigenvalue problems was already proposed by Rutishauser in 1954 with an infinitesimal version of the QR algorithm \cite{Rutishauser1954}. The Toda flow is another famous flow known from the study of the Toda lattice in statistical mechanics \cite{Toda1967}. Its relation to a matrix flow for tridiagonal matrices was understood by Flachka and Moser in 1974 and 1975 \cite{Flaschka1974a,Moser}. This flow was generalized and applied to banded matrices by Mielke 1998 \cite{Mielke1998}. Stein was one of the first to solve continuous unitary transformations of that flow perturbatively in 1997 \cite{Stein1997} and the flow was generalized further by Knetter and Uhrig in 2000, where they introduced the quasi-particle generator $\eta_{\mathrm{QP}}$ \cite{Knetter2000}. They obtained a general perturbative solution for this flow equation under the special condition of equidistant spectrum of $\mathcal{H}_0$ and called it perturbative continuous unitary transformations (pCUT). In an eigenbasis of $\mathcal{H}_0$ the quasi-particle generator $\eta_{\mathrm{QP}}$ can be defined as 
\begin{equation}
	\eta_{\mathrm{QP},i,j}(l) = \mathrm{sgn} (\mathcal{H}_{0,i,i}-\mathcal{H}_{0,j,j}) \mathcal{H}_\mathrm{i,j}(l).
	\label{eq::QP generator}
\end{equation}
By stating $\mathcal{H}(0)$ is linked we define what processes are considered as linked. The off-diagonal parts of $\mathcal{H}(0)$ are assumed to be local operators. Two local operators commute when they act on disconnected parts of the lattice. As $\eta_{\mathrm{QP}}(0)$ decouples all blocks of $\mathcal{H}(0)$, it is also linked and can be written as a sum of local operators. Then by definition of the flow equation~\eqref{eq::flow equation}, the cluster-additivity property is ensured during the flow as the commutator vanishes for local operators acting on disconnected clusters.\\
The second genuinely linked-cluster transformation is the multi-block orthogonalization method (MBOT) \cite{Trebst2000,Zheng2001}. A similar construction can also be found in \cite{Datta1996}. As the name indicates, also here it is crucial that all blocks of the Hamiltonian are decoupled. This transformation is constructed with the matrix exponential and a global generator $\mathcal{S}$, i.e. $T=\exp(-\mathcal{S})$. It makes use of the connection between Lie algebra and matrix exponential as well as the linked structure established by the commutator expansion
\begin{equation}
		\exp(\mathcal{S})\mathcal{H}\exp(-\mathcal{S}) = \sum_{n=0}^\infty \frac{[(\mathcal{S})^n,\mathcal{H}]}{n!},\quad \text{where} \quad  [(\mathcal{S})^n,\mathcal{H}] \equiv [\underbrace{\mathcal{S},\dots [\mathcal{S},[\mathcal{S}}_{\textit{n \rm{times}}},\mathcal{H}]]\dots].
\label{eq::Lie algebra expansion matrix exponential}
\end{equation}
It is constructed order by order demanding that up to a given order all off-diagonal elements between different blocks of $\mathcal{H}_\mathrm{eff}$ vanish. As the first order part of $X$ has to decouple all blocks, it can be written as a sum of local operators. From the form of \eqref{eq::Lie algebra expansion matrix exponential} it is then ensured that the transformation is linked cluster in the next order if $\mathcal{S}$ contains only linked-terms in all previous orders. For the sake of completeness we mention that in \cite{Datta1996} also a local transformation constructed order by order as 
\begin{equation}
	T=\exp(-\lambda\mathcal{S}_1)\cdot \ldots \cdot \exp(-\lambda^n\mathcal{S}_n)
	\label{eq::transformation as product of exponentials from Datta1996}
\end{equation}
is introduced.\\ 
Both pCUT and MBOT can be constructed order by order in a model-independent form for Hamiltonians with equidistant $\mathcal{H}_0$. There is also a model-dependent method to use $\eta_{\mathrm{QP}}$ perturbatively (epCUT) and non-perturbatively (deepCUT) \cite{Krull2012a} for $\mathcal{H}_0$ with non-equidistant spectrum directly in the thermodynamic limit. Also, recently an extension of the pCUT approach to multiple quasiparticle types as well as non-Hermitian Hamiltonians and open systems was introduced under the name $\mathrm{pcst}^{\texttt{++}}$ \cite{Lenke2023}. It should also be possible to write down model-independent perturbative expressions for MBOT and $\mathcal{H}_0$ with non-equidistant spectrum similarly as in the Schrieffer-Wolff expansion of the minimal transformation but now using projectors on all eigenspaces of $\mathcal{H}_0$.  Unfortunately, it is hard to transfer the MBOT method to non-perturbative exact calculations on finite graphs since it is difficult to find a transformation that sets all block-diagonal parts of $\mathcal{S}$ to zero while block-diagonalising the Hamiltonian. Also how to efficiently truncate the basis states for MBOT is not clear non-perturbatively. In contrast, the application of flow equations using $\eta_{\mathrm{QP}}$ to non-perturbative problems on finite systems is straightforward and was used in the gCUT approach \cite{Yang2011}. With regards to basis truncations it is important to realize that one can use a modified version of the generator $\eta_{\mathrm{QP}}$ 
\begin{equation}
	\eta_{\mathrm{QP},i,j}^n(l) = \left(1-\Theta(\mathcal{H}_{0,i,i}-e_0^{n+1})\, \Theta(\mathcal{H}_{0,j,j}-e_0^{n+1})\right)\mathrm{sgn} (\mathcal{H}_{0,i,i}-\mathcal{H}_{0,j,j})  \mathcal{H}_\mathrm{i,j}(l)
	\label{eq::modified QP generator}
\end{equation}
and still obtain the same effective Hamiltonian in the blocks $m\leq n$ \cite{Fischer2010}. To see this we introduce the set of indices in the $n$-particle block $s_n$. Then we note that the special form of $\eta_{\mathrm{QP}}$ leaves the flow in lower subspaces $m\leq n$ invariant under unitary transformations of the higher subspaces $m>n$ as can be seen by 
\begin{equation}
	\sum_k \mathcal{H}_{i,k}(l)\mathcal{H}_{k,j}(l)= \sum_k (\mathcal{H}U_{i,k})(l)(U^\dagger\mathcal{H}(l))_{k,j}
	\label{eq::unitary invariance eta_QP}
\end{equation}
with $i,j$ in the subspaces $\bigcup_{m\leq n}s_m$ and $k$ in the higher-energy spaces $\bigcup_{m > n}s_m$ and $U$ a unitary matrix acting on the states $k$. As a consequence, one can efficiently truncate the basis states using the Krylov subspace of $\bigoplus_{m=0}^n\mathscr{H}_0^m$ when targeting the subspace $n$ of $\mathcal{H}_\mathrm{eff}$ with the quasi-particle generator because states of higher orders of the Krylov subspace only contribute at larger times $l$ of the flow. This efficient way of truncating is a big advantage of the special form of $\eta_{\mathrm{QP}}$ and distinguishes this generator. With this we conclude the discussion of existing cluster-additive block-diagonalisation methods.

\subsection{Projective block-diagonalisation method}
\label{subsection::Projective block diagonalisation}
Another type of transformation is the projective transformation $T$ constructed of the eigenstates and energies of the block $n$ of interest. This transformation can be given in an order-independent form, needs minimal information to be constructed, has minimal norm $\lVert \mathds{1} - T \rVert$ and in many situations can be implemented numerically more efficiently than the previous transformations because only matrix-vector multiplications are needed and for most cases obtaining energies and eigenstates with Krylov-based algorithms is faster than solving differential equations. Unfortunately, it only allows for a linked-cluster expansion of excitations under special circumstances.\\
The projective transformation is constructed by projectors $P_n$ on the eigenspaces of $\mathcal{H}_0$ and projectors $\bar{P}_n$ on the adiabatically connected eigenspaces of $\mathcal{H}_\mathrm{eff}$. Projectors are idempotent operators, i.e. $P_n^2 = P_n$ and $\bar{P}_n^2 = \bar{P}_n$. For $v \in \mathscr{H}$ 
\begin{equation}
	P_n \, v \in \mathscr{H}_0^n 
	\label{eq::projector P_n}
\end{equation}
and 
\begin{equation}
	\bar{P}_n \,v \in \mathscr{H}_{\mathrm{eff}}^n.
	\label{eq::projector bar P_n}
\end{equation}
Further, from the orthogonality of the subspaces the resolution of identity 
\begin{equation}
	\mathds{1} = \sum_n P_n = \sum_n \bar{P}_n
	\label{eq::resolution of identity projector}
\end{equation}
follows. A good educational introduction to perturbation theory described in the framework of projection operators is given in \cite{Yao2000}.\\
We first state the form of the projective transformation introduced by Takahashi \cite{Hasegawa1977}:
\begin{equation}
	T=\sum_n T_n
	\label{eq::Takahashi transformation for all blocks}
\end{equation}
\begin{equation}
	T_n=\bar{P}_n P_n \left(\sum_m P_m\bar{P}_m P_m\right)^{-1/2}
	\label{eq::Takahashi transformation for one block}
\end{equation}
He further used a result of Kato \cite{Kato1949} for the perturbative form of the projector $\bar{P}_n$
\begin{equation}
	\bar{P}_n = P_n - \sum_{s=1}^{\infty} \quad \sum_{k_1+\dots+k_{s+1}=s,\,\, k_i\leq 0} S_n^{k_1}VS_n^{k_2}V\dots VS_n^{k_{s+1}},
	\label{eq::Kato perturbative expansion of projector on H subspaces}
\end{equation}
where $S_n^0 \equiv -P_n, \quad S_n^k\equiv \left(\frac{1-P_n}{e_0^n-\mathcal{H}_0}\right)^k$
and realized that $P_n\left(\sum_mP_m\bar{P}_m P_m\right)^{-1/2}P_n$ can be expanded similarly using Kato's expression. Note that while $P_n\bar{P}_nP_n$ can not be inverted its restriction to the subspace $\mathscr{H}_0^n$ can. The local expressibility of the transformation is important as it shows that the transformation has no contributions on subgraphs of the lattice with a larger number of bonds than the perturbation order. The transformation $T$ is symmetric in the diagonal blocks as can be seen by 
\begin{equation}
	P_nTP_n = P_nT_nP_n = P_n\bar{P}_n P_n \left(\sum_m P_m\bar{P}_m P_m\right)^{-1/2} P_n= P_n\left(\sum_m P_m\bar{P}_m P_m\right)^{1/2}P_n
	\label{eq::showing symmetry in diagonal block for Takahahsi part one}
\end{equation}
and 
\begin{equation}
	P_n^{\phantom\dagger}T^\dagger P_n^{\phantom\dagger} = P_n^{\phantom\dagger}T_n^\dagger P_n^{\phantom\dagger} = P_n \left(\sum_m P_m\bar{P}_m P_m\right)^{-1/2}  P_n\bar{P}_n P_n = P_n \left(\sum_m P_m\bar{P}_m P_m\right)^{1/2} P_n.
	\label{eq::showing symmetry in diagonal block for Takahahsi part two}
\end{equation}
This shows equivalence of the perturbative expansion of $T$ with the two-block orthogonalization method (TBOT) \cite{Zheng2006} as for TBOT in \cite{Zheng2006} it was shown that any perturbative transformation that decouples two blocks of the Hamiltonian is uniquely determined by demanding symmetric diagonal blocks.\\
 The projective transformation can also be written in the form of a Schrieffer-Wolff transformation $T_{\mathrm{SW}}=\exp(-\mathcal{S}_{\mathrm{SW}})$ that decouples block $n$ from the rest. We understand as a Schrieffer-Wolff transformation $T_{\mathrm{SW}}$ any transformation with a particular anti-block-diagonal form of $\mathcal{S}_{\mathrm{SW}}$. Introducing
\begin{equation}
	R=\sum_{m,\,m\neq n} P_m
	\label{eq::projector on all subspaces except n}
\end{equation}
it can be written as 
\begin{equation}
	T_{\mathrm{SW}} = \left(\bar{P}_n P_n +\bar{R}R\right) \left(P_n\bar{P}_n P_n + R\bar{R} R \right)^{-1/2} = \exp(-\mathcal{S}_{\mathrm{SW}}),
	\label{eq::Schrieffer-Wolff form of projective transformation}
\end{equation}
where $\mathcal{S}_{\mathrm{SW}}$ takes the form
\begin{equation}
	\mathcal{S}_{\mathrm{SW}}	=\begin{pmatrix}
		0 & \mathcal{S}_{\mathrm{SW},n,R} \\
		-\mathcal{S}_{\mathrm{SW},n,R}^\dagger & 0 \\
	\end{pmatrix}.
	\label{eq::Schrieffer-Wolff global generator form of projective transformation}
\end{equation}
That $\mathcal{S}_{\mathrm{SW}}$ has to take such a form follows at least perturbatively from the uniqueness of $\mathcal{S}_{\mathrm{SW}}$, the symmetry of $T_{\mathrm{SW}}$ in its diagonal blocks, and the fact that an exponential of an anti-block diagonal $\mathcal{S}_{\mathrm{SW}}$ as in Eq.~\eqref{eq::Schrieffer-Wolff global generator form of projective transformation} yields a transformation that is symmetric in the diagonal blocks. In \cite{Shavitt1980} the transformation is constructed perturbatively by an $\mathcal{S}_{\mathrm{SW}}$ of that form and it is called canonical form of van Vleck perturbation theory. A review of the Schrieffer-Wolff transformation also constructs the transformation order by order this way \cite{Bravyi2011}, while also giving a very convenient form of the transformation as direct rotation 
\begin{equation}
	T_{\mathrm{SW}} = \sqrt{(\bar{P}_n-\bar{R})(P_n-R)}
	\label{eq::Schrieffer-Wolff direct rotation form}
\end{equation}
between $P_n$ and $\bar{P}_n$, i.e.
\begin{equation}
	T_{\mathrm{SW}}^\dagger \bar{P}_n^{\phantom\dagger} T_{\mathrm{SW}}^{\phantom\dagger} = P_n^{\phantom\dagger}.
	\label{eq::Schrieffer-Wolff connection between P_n and bar P_n}
\end{equation}
The equivalence between \eqref{eq::Schrieffer-Wolff form of projective transformation} and \eqref{eq::Schrieffer-Wolff direct rotation form} is most easily seen by comparing 
\begin{equation}
	\left(\bar{P}_n P_n +\bar{R}R\right)^2 = \bar{P}_n P_n\bar{P}_n P_n + \bar{R}R\bar{R}R + \bar{P}_n P_n\bar{R}R + \bar{R}R\bar{P}_n P_n
	\label{eq::square lhs Schrieffer-Wolff form of projective transformation}
\end{equation}
and 
\begin{equation}
	(\bar{P}_n-\bar{R})(P_n-R)\left(P_n\bar{P}_n P_n + R\bar{R} R\right) = \bar{P}_n P_n\bar{P}_n P_n + \bar{R}R\bar{R}R - \bar{P}_nR\bar{R}R - \bar{R}P_n\bar{P}_nP_n.
	\label{eq::product square root argument of Schrieffer-Wolff direct rotation form and rhs of Schrieffer-Wolff form of projective transformation}
\end{equation}
The expressions are identical since $\mathds{1} = P_n + R$ and $\bar{P}_n\bar{R} = 0$. In \cite{Bravyi2011} the transformation is constructed perturbatively order by order using the form of the matrix exponential Eq.~\eqref{eq::Schrieffer-Wolff global generator form of projective transformation}. This is not necessary as Takahashi's form of the transformation for the effective low-energy block is exactly identical and can be written down non-inductively. Another unique property of $T_{\mathrm{SW}}$ is that it has minimal norm $\lVert \mathds{1} - T_{\mathrm{SW}} \rVert$ of all possible transformations that decouple the block $n$ from the rest \cite{Davis1969,Bravyi2011}. In contrast to the MBOT transformation, the global generator only is anti-block-diagonal with respect to two blocks and because of that has non-local anti-block-diagonal terms in general.\\
At last we state the form of the transformation given in \cite{Cederbaum1989}. It is very similar to Takahashi's form but given in terms of eigenvectors instead of projectors. This form will be particularly useful for the construction of the cluster-additive projective transformation in Sec.~\ref{section::Projective transformation fulfilling the linked-cluster property}. The eigenvectors and energies $X_0$ and $D_0$ of $\mathcal{H}_0$ and $X$ and $D$ of $\mathcal{H}$ fulfil
\begin{equation}
	\mathcal{H} X_0 = X_0 D_0
	\label{eq::eigendecomposition H_0}	
\end{equation}
and
\begin{equation}
	\mathcal{H} X = X D.
	\label{eq::eigendecomposition H}
\end{equation}
Projection operators and eigenvectors are related by 
\begin{equation}
	P_{n,i,j} = \sum_{k\in s_n} X_{0,i,k}^{\phantom\dagger}X^\dagger_{0,k,j}
	\label{eq::P_n_bar relation to eigenvectors of H}
\end{equation}
and
\begin{equation}
	\bar{P}_{n,i,j} = \sum_{k\in s_n} X_{i,k}^{\phantom\dagger}X^\dagger_{k,j},
	\label{eq::P_n relation to eigenvectors of H_0}	
\end{equation}
where the ordering of basis states and energies is such that $X_{0,i,j}$ is only non-zero for $i,j\in s_n$. Here we remind that the set of indices in the $n$-particle block is denoted by $s_n$. Introducing 
\begin{equation}
	X^{P_n} \equiv P_n X P_n 
	\label{eq::H_0 block n of eigenvectors}
\end{equation}
one can then write the transformation as 
\begin{equation}
	T_{n,i,j}= \sum_k X_{i,k}^{\phantom\dagger} \left(X^{P_n\,\dagger} \left(\sum_m X^{P_m\phantom \dagger}X^{P_m\,\dagger}\right)^{-1/2}\right)_{k,j}
	\label{eq::projective transformation as given in Cederbaum et al.}
\end{equation}
with $k\in s_n$. In \cite{Cederbaum1989} it was proved that this transformation has minimal norm $\lVert \mathds{1} - T \rVert$, which shows that also when one wants to decouple all blocks and not just two as in $T_{\mathrm{SW}}$ this is the transformation with minimal norm. The MBOT method, which is a Schrieffer-Wolff transformation of local anti-block-diagonal operators, is different and consequently does not have minimal norm. Hence, only when one decouples two blocks an anti-block-diagonal $\mathcal{S}_{\mathrm{SW}}$ leads to a transformation with minimal norm $\lVert \mathds{1} - T_{\mathrm{SW}} \rVert$.\\
For the effective Hamiltonian in the desired block $n$ only the part $X^{P_n\phantom \dagger}X^{P_n\,\dagger}$ contributes. By denoting the restriction of $X^{P_n}$ to the basis states $s_n$ with $X_{s_n}^{P_n}$ the part of the transformation that creates the effective Hamiltonian in block $n$ can be written as
\begin{equation}
	T_{n,i,s_n}= \sum_{k\in s_n} X_{i,k}^{\phantom\dagger} \left( X_{s_n}^{P_n\,\dagger} \left(X_{s_n}^{P_n\phantom \dagger}X_{s_n}^{P_n\,\dagger}\right)^{-1/2}\right)_{k,\,s_n}.
	\label{eq::projective transformation only creating effective Hamiltonian in block n as given in Cederbaum et al.}
\end{equation}
As these are the only basis states for which $X^{P_n}$ has non-zero matrix elements this restricts the transformation to the relevant part for each block and can help making considerations easier. In particular, for two disconnected clusters $A$ and $B$ and transformations $T_{l,A}$ in $A$ and $T_{k,B}$ in $B$ and a transformation $T_{n,s_l\otimes s_k}$ on $A\cup B$ in the subspace $n$, where \mbox{$e_0^n-e_0^0=(e_0^l-e_0^0)+(e_0^k-e_0^0)$}, that projects only on the states $s_l\otimes s_k$ (but only on this, not on the whole block $n$ on $A \cup B$) one finds 
\begin{equation}
	H_{\mathrm{eff},s_l\otimes s_k}(A\cup B) = H_{\mathrm{eff},s_l}(A) \otimes \mathds{1}_B + \mathds{1}_A \otimes H_{\mathrm{eff},s_k}(B)
	\label{eq::additivity effective Hamiltonian for tensor product of subspaces}
\end{equation}
as 
\begin{equation}
	\sum_{i,j}X_{s_l\otimes s_k,i}^\dagger\mathcal{H}_{i,j}X_{j,s_l\otimes s_k}^{\phantom\dagger} = D_{s_l}(A)\otimes \mathds{1}_B + \mathds{1}_A\otimes D_{s_k}(B)
	\label{eq::cluster-additivity eigenvalues s_l, s_k and s_l otimes s_k}
\end{equation}
and 
\begin{equation}
	\left( X_{s_l\otimes s_k}^{P_n\,\dagger} \left(X_{s_l\otimes s_k}^{P_n}X_{s_l\otimes s_k}^{P_n\,\dagger}\right)^{-1/2}\right) = \left( X_{s_l}^{P_l\,\dagger} \left(X_{s_l}^{P_l\phantom \dagger}X_{s_l}^{P_l\,\dagger}\right)^{-1/2}\right) \otimes \left( X_{s_k}^{P_k\,\dagger} \left(X_{s_k}^{P_k\phantom \dagger}X_{s_k}^{P_k\,\dagger}\right)^{-1/2}\right).
	\label{eq::transformation in s_l otimes s_k as tensor product of transformations in s_l and s_k}
\end{equation}
This was also shown in \cite{Bravyi2011} and shows that the effective Hamiltonian of the projective transformation allows to perform a linked-cluster decomposition for degenerate ground states. For excitations it is not helpful since one can not separate excitations in $A\cup B$ with one excitation in $A$ and ground state in $B$ from ground state in $A$ and one excitation in $B$. The problems caused by this will become obvious in the next subsection, where we show the failure of a linked-cluster expansion for spin-flip excitations in a simple toy model.

\subsection{Failure of linked-cluster expansion for excited states with projective method}
\label{subsection::Failure of linked-cluster expansion for excited states with projective method}
\begin{figure}[!htbp]
	\centering	
	\includegraphics[width=0.8\textwidth]{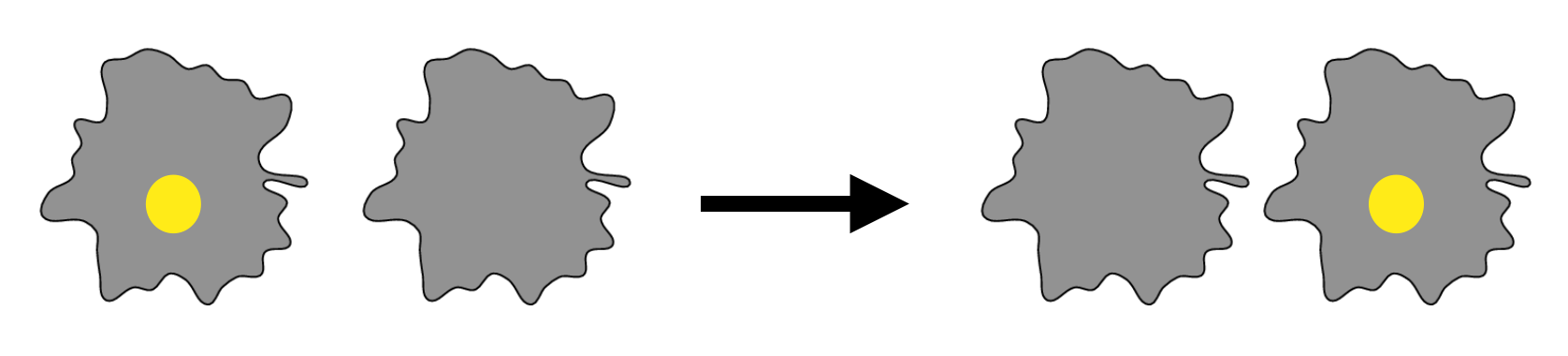}
	\caption[Hopping between two disconnected clusters]{The figure depicts a hopping process of one particle (yellow ball) between two disconnected clusters. For the Hamiltonian \eqref{eq::toy Hamiltonian linked-cluster violation projective transformation} such hopping elements are seen in the effective one-particle Hamiltonian in order four of perturbation. These processes are a manifestation of the violation of cluster-additivity of the minimal projective transformation.}
	\label{fig::Hopping between two disconnected clusters}
\end{figure}

Gelfand realized that a linked-cluster expansion for elementary excitations is possible with non-cluster additive transformations as long as the elementary excitations have a different quantum number than the ground state \cite{Gelfand1996}. To show the failure of a linked-cluster expansion for the minimal transformation we therefore consider a high-field expansion of the Hamiltonian given as the sum of the transverse-field Ising chain, where this is given, and a parity breaking term $\sigma_{z,\nu}\sigma_{x,\nu+1}$:
\begin{equation}
	\mathcal{H} = \sum_\nu \sigma^{z}_\nu + \sum_\nu  \left(\lambda\sigma^{x}_\nu\sigma^{x}_{\nu+1}+\mu\left(\sigma^{z}_\nu\sigma^{x}_{\nu+1}+\sigma^{x}_\nu\sigma^{z}_{\nu+1}\right)\right)
	\label{eq::toy Hamiltonian linked-cluster violation projective transformation}
\end{equation}
The Pauli matrices $\sigma^{x/z}_\nu$ describe spins-1/2 on site $\nu$. For $\mu\neq 0$ ground state and spin-flip excitations are coupled to each other. Now we consider two disconnected clusters $A$ and $B$. The Hamiltonian on $A \cup B$ can be written as 

\begin{equation}
	\mathcal{H} = \mathcal{H}_A + \mathcal{H}_B,
	\label{eq::H as sum of H_A and H_B}
\end{equation}
where
\begin{equation}
	[\mathcal{H}_A,\mathcal{H}_B] = 0
	\label{eq::H_A and H_B commute}
\end{equation}
holds. Consequently the eigenfunctions of $H_{A\cup B}$ take the form
\begin{equation}
	\ket{\Psi}_{A\cup B} = \ket{\Psi}_A \otimes \ket{\Psi}_B
	\label{eq::form of eigenfunctions on A cup B}
\end{equation}
and have an energy 
\begin{equation}
	\mathcal{H} \ket{\Psi} = (\mathcal{H}_A\ket{\Psi}_A) \otimes \ket{\Psi}_B + \ket{\Psi}_A \otimes (\mathcal{H}_A\ket{\Psi}_B) = (e_A+e_B) \ket{\Psi}.
	\label{eq::energy of eigenfunctions on A cup B}
\end{equation}
For spin-flip excitations on $A\cup B$ it follows that they are either build of a ground state on $A$ and a spin-flip excitation on $B$ or vice versa:
\begin{equation}
	\ket{\Psi}_{1,\,A\cup B} = \ket{\Psi}_{1,\,A} \otimes \ket{\Psi}_{0,\,B} \quad \lor \quad  \ket{\Psi}_{1,\,A\cup B} = \ket{\Psi}_{0,\,A} \otimes \ket{\Psi}_{1,\,B}
	\label{eq::form of spin-flip excitation on A cup B}
\end{equation}
For the case $\mu=0$ where the parity is not broken, $P_0\ket{\Psi}_1=0$. Then $X_{s_1}^{P_1}$ is block-diagonal in the $A$- and $B$-blocks
\begin{equation}
	X_{s_1}^{P_1} = \begin{pmatrix}
		X_{s_1,A}^{P_1}X_{s_0,B}^{P_0} & 0 \\
		0 & X_{s_1,B}^{P_1}X_{s_0,A}^{P_0} \\
	\end{pmatrix}
\label{eq::X_1^s_1 for mu=0}
\end{equation}
and additivity of $\bar{\mathcal{H}}_{\mathrm{eff}}^1$ is given
\begin{equation}
	T_1^\dagger \mathcal{H} T_1 - e^0(A\cup B) = \bar{\mathcal{H}}_{\mathrm{eff}}^1(A\cup B) = \bar{\mathcal{H}}_{\mathrm{eff}}^1(A) \oplus \bar{\mathcal{H}}_{\mathrm{eff}}^1(B).
\end{equation}
This is not the case when $\mu\neq 0$. Then $P_0 \ket{\Psi}_1\neq 0$ and $X_{s_1}^{P_1}$ is not block-diagonal in the $A$- and $B$-blocks any more
\begin{equation}
	X_{s_1}^{P_1} = \begin{pmatrix}
		X_{s_1,A}^{P_1}X_{s_0,B}^{P_0} & X_{s_1,A}^{P_0}X_{s_0,B}^{P_1} \\&\\
		X_{s_1,B}^{P_0}X_{s_0,A}^{P_1} & X_{s_1,B}^{P_1}X_{s_0,A}^{P_0} \\
	\end{pmatrix}.
	\label{eq::X_1^s_1 for mu not equal 0}
\end{equation}
Consequently, additivity of $\bar{\mathcal{H}}_{\mathrm{eff}}^1$
\begin{equation}
	T_1^\dagger \mathcal{H} T_1 - e^0(A\cup B) = \bar{\mathcal{H}}_{\mathrm{eff}}^1(A\cup B) \neq \bar{\mathcal{H}}_{\mathrm{eff}}^1(A) \oplus \bar{\mathcal{H}}_{\mathrm{eff}}^1(B).
	\label{eq::toy model mu not equal 0 additivity not given any more}
\end{equation}
is not given any more. If one performs calculations for the model with $\mu=1$ one finds these non-linked terms in order four. Particles can then hop between disconnected clusters as illustrated in Fig.~\ref{fig::Hopping between two disconnected clusters}, which is never allowed in a linked-cluster expansion. The crucial step for the construction of a cluster additive projective transformation is to modify $X_{s_1}^{P_1}$ to restore block-diagonal form for the general case $\mu\neq 0$ and to eliminate these hopping elements between disconnected clusters.
\section{Projective cluster-additive transformation}
\label{section::Projective transformation fulfilling the linked-cluster property}

In the last section we reviewed the minimal projective transformation and showed an example where the failure of linked-cluster expansion for excited states was shown. In particular, the problem could be seen in the non-block diagonal form of $X_{s_1}^{P_1}$ in \eqref{eq::X_1^s_1 for mu not equal 0}. It is the major achievement of this paper to introduce the projective cluster-additive transformation $T^{\rm pca}$ which cures this problem.

\subsection{Cluster-additivity for single particle states}

It is necessary to modify $X_{s_1}^{P_1}$ to $\tilde{X}_{s_1}^{P_1}$ to obtain a cluster-additive transformation for single-particle states. To achieve this we modify the eigenstates of $\mathcal{H}$. For ground-state energies additivity is always given and consequently the ground state $\ket{\Psi}_{0}$ is not modified:
\begin{equation}
	\ket{\tilde{\Psi}}_{0} = \ket{\Psi}_{0} 
	\label{eq::lc-transformed ground state}
\end{equation}
For single-particle eigenstates $\ket{\Psi}_{1}$ we modify in the following way,
\begin{equation}
	\ket{\tilde{\Psi}}_{1} = \ket{\Psi}_{1} -  \left(1/\langle 0 \vert \Psi_{0} \rangle \right)\langle 0 \vert \Psi_{1} \rangle\ket{\Psi}_{0},
	\label{eq::lc-transformed single particle states}
\end{equation}
where $\vert 0 \rangle$ denotes the unperturbed ground state. Note that in general the states $\ket{\tilde{\Psi}}_{1}$ as well as $\ket{\tilde{\Psi}}_{0}$ and $\ket{\tilde{\Psi}}_{1}$ are not orthogonal and normalized any more. The ground-state subtraction of $\ket{\Psi}_{0}$ in $\ket{\tilde{\Psi}}_{1}$ leads to
\begin{equation}
	P_0\ket{\tilde{\Psi}}_{1} = 0.
	\label{eq::P_0 tildePsi_1 = 0}
\end{equation}
As long as $\langle 0 \vert \Psi_{0} \rangle \neq 0$ this subtraction is unique. Recalling the form \eqref{eq::form of spin-flip excitation on A cup B} of a single-particle eigenstate on two disconnected clusters $A\cup B$ we find
\begin{equation}
	\ket{\tilde{\Psi}}_{1,\,A\cup B} = \ket{\tilde{\Psi}}_{1,\,A} \otimes \ket{\tilde{\Psi}}_{0,\,B}.
	\label{eq::cluster additivity ground-state subtraction}
\end{equation}
$\tilde{X}_{s_1}^{P_1}$ then takes the form 
\begin{equation}
	\tilde{X}_{s_1}^{P_1} = \begin{pmatrix}
		\tilde{X}_{s_1,A}^{P_1}\tilde{X}_{s_0,B}^{P_0} & 0 \\
		0 & \tilde{X}_{s_1,B}^{P_1}\tilde{X}_{s_0,A}^{P_0} \\
	\end{pmatrix} 
    \label{eq::block-form of tildeX_1}
\end{equation}
because $\tilde{X}_{s_1,A}^{P_0}=\tilde{X}_{s_1,B}^{P_0}=0$. The linked-cluster transformation of the single-particle block can now be conveniently written as  
\begin{equation}
	T_{1,i,s_1}^{\mathrm{pca}}= \sum_{k\in s_1}X_{i,k} \left(\tilde{X}_{s_1}^{P_1\,\dagger} \left(\tilde{X}_{s_1}^{P_1\phantom \dagger}\tilde{X}_{s_1}^{P_1\,\dagger}\right)^{-1/2}\right)_{k,s_1}.
	\label{eq::linked-cluster transformation single-particle block}
\end{equation}
Particularly important is the part
\begin{equation}
	\mathcal{T}_{1,s_1,s_1}^{\mathrm{pca}}= \left(\tilde{X}_{s_1}^{P_1\,\dagger} \left(\tilde{X}_{s_1}^{P_1\phantom \dagger}\tilde{X}_{s_1}^{P_1\,\dagger}\right)^{-1/2}\right)_{s_1,s_1}
	\label{eq::projected eigenvector part of single-particle linked-cluster transformation}
\end{equation}
since its form determines the matrix elements of $\mathcal{H}_{\mathrm{eff}}^n$. As we have already seen, this part is block-diagonal
\begin{equation}
	\mathcal{T}_{1,A\cup B}^{\mathrm{pca}} = \mathcal{T}_{1,A}^{\mathrm{pca}} \oplus \mathcal{T}_{1,B}^{\mathrm{pca}}.
	\label{eq::block-diagonality mathcal T}
\end{equation}
The other part of the transformation just yields a diagonal matrix 
\begin{equation}
	\sum_{i,j}X^\dagger_{s_1,i}\mathcal{H}_{i,j} X_{j,s_1}^{\phantom\dagger} = D_A \oplus D_B.
	\label{eq::diagonal values single-particle block}
\end{equation}
Combining the direct sum of eigenvalues on $A\cup B$
\begin{equation}
	D_A \oplus D_B - e^0(A\cup B) = e^1_A \oplus e^1_B
	\label{eq::direct sum of eigenvalues on A cup B}
\end{equation}
with the form of $\mathcal{T}_1^{\mathrm{pca}}$ in Eq.~\eqref{eq::block-diagonality mathcal T} one obtains additivity of $\bar{\mathcal{H}}_{\mathrm{eff}}^1$:
\begin{equation}
	\sum_{r,k} T_{1,s_1,r}^{\mathrm{pca},\dagger} \mathcal{H}_{r,k} T_{1,k,s_1}^{\mathrm{pca}} - e^0(A\cup B) = \bar{\mathcal{H}}_{\mathrm{eff}}^1(A\cup B) = \bar{\mathcal{H}}_{\mathrm{eff}}^1(A) \oplus \bar{\mathcal{H}}_{\mathrm{eff}}^1(B)
	\label{eq::cluster-additivity T^lc for single-particle excitations}
\end{equation}
For one-particle excitations we now already have constructed the right transformation. The more general case of multi-particle excitations will be discussed in the next subsection.

\subsection{Cluster-additivity for multi-particle excitations}

As mentioned before, the cluster additivity of the effective Hamiltonian implies that we can construct additive irreducible operators in every block of interest of the effective Hamiltonian. To show cluster-additivity for multi-particle excitations we again make use of the tensor product structure of eigenstates on $A\cup B$ with $A$ and $B$ not connected for $n$-particle states $\ket{\Psi}_{n}$ with energy $e_{0,\,A}^a+e_{0,\,B}^b=e_{0,\,A\cup B}^n$ of $\mathcal{H}_0$:
\begin{equation}
	\ket{\Psi}_{n,\,A\cup B} = \ket{\Psi}_{a,\,A} \otimes \ket{\Psi}_{b,\,B}\, .
	\label{eq::tensor-product structure multi-particle excitation on A cup B}
\end{equation}
What changes compared to single-particle excitations is the transformation of eigenstates $\ket{\Psi}\rightarrow \ket{\tilde{\Psi}}$ for the construction of the transformation. For a state with energy $e_0^n$ we demand that the projection on eigenstates of $\mathcal{H}_0$ with $e_0^m<e_0^n$ is zero, i.e. for
\begin{equation}
	R=\sum_{m,\,m<n} P_m
	\label{eq::projector on lower-energy eigenspaces}
\end{equation}
we need to have
\begin{equation}
	R\ket{\tilde{\Psi}}_n = 0.
	\label{eq::low-energy part of eigenstate is zero}
\end{equation}
This has to be achieved by subtracting lower-energy eigenstates of $\ket{\tilde{\Psi}}_n$. As long as 
\begin{equation}
	Y_{n-1}=X_{i,j} \quad , \quad i,j \in \cup_{m<n}s_m \, , 
	\label{eq::matrix that has to be inverted for multi-particle excitation}
\end{equation}
is invertible the construction is always possible and unique. Assuming non-singular $Y_{n-1}$, the transformed states $\ket{\tilde{\Psi}}_n$ are defined as 
\begin{equation}
	\ket{\tilde{\Psi}}_n = \ket{\Psi}_n - \sum_{m<n} \left[Y_{n-1}^{-1}\left(R\ket{\Psi}_n\right)\right]_m\Ket{\Psi}_{m}.
	\label{eq::transformed tilde psi states}
\end{equation} 
The singular values of $Y_{n-1}$ are the square roots of the eigenvalues of 
\begin{equation}
	W_{n-1}=\sum_{m<n}P_m \sum_{m<n}\bar{P}_m  \sum_{m<n}P_m.
    \label{eq::particle decay characteristic generalized}
\end{equation} 
As we discuss later in the context of NLCEs (see Subsec.~\ref{subsection::Non-perturbative results for single spin flip and bound states}), particle-decay highly influences the convergence properties of the expansion. For particle-decay of $n$-particle states it is important to investigate the behaviour of $W_n$ and not of $W_{n-1}$. The reason is that particle-decay of the $n$-particle states would show up as a problem in the construction of $m$-particle states with $m>n$. When the smallest eigenvalue of $W_{n}$ drops to almost zero sharply, this is a hallmark of particle-decay.
The transformation from $\ket{\Psi}_n$ to $\ket{\tilde{\Psi}}_n$ can be visualized as
\begin{equation}
	\begin{pmatrix}
		{\color{blue}P_0 \ket{\Psi}_n}\\
		{\color{blue}\vdots} \\
		{\color{red}P_{n} \ket{\Psi}_n}\\
		{\color{red}\vdots} \\
		{\color{red}P_{N} \ket{\Psi}_n}
	\end{pmatrix} \rightarrow 
 \begin{pmatrix}
	{\color{blue}0}               \\
	{\color{blue}\vdots} \\
	{\color{red}P_{n} \ket{\tilde{\Psi}}_n}\\
	{\color{red}\vdots} \\
	{\color{red}P_{N} \ket{\tilde{\Psi}}_n}
\end{pmatrix}.
\label{eq::T from Psi to tilde Psi}
\end{equation}
Since this subtraction is unique for non-singular $Y_{N-1}$ in Eq.~\eqref{eq::matrix that has to be inverted for multi-particle excitation}, it follows
\begin{equation}
	\ket{\tilde{\Psi}}_{n,\,A\cup B} = \ket{\tilde{\Psi}}_{a,\,A} \otimes \ket{\tilde{\Psi}}_{b,\,B}\, .
	\label{eq::cluster-additivity transformed tilde multi-particle excitations}
\end{equation}
Eq.~\eqref{eq::cluster-additivity transformed tilde multi-particle excitations} is at the heart of the cluster-additivity of the transformation. It follows
\begin{equation}
	\tilde{X}_{s_a\otimes s_b}^{P_n} = \tilde{X}^{P_a}_{s_a,\,A} \otimes \tilde{X}^{P_b}_{s_b,\,B}
	\label{eq::tensor-product of tilde X on A and B for multi-particle excitation}
\end{equation}
and with that for the transformation
\begin{equation}
	\tilde{X}_{s_a\otimes s_b}^{P_n\,\dagger} \left(\tilde{X}_{s_a\otimes s_b}^{P_n\phantom \dagger}\tilde{X}_{s_a\otimes s_b}^{P_n\,\dagger}\right)^{-1/2} = \tilde{X}_{s_a,\,A}^{P_a\,\dagger} \left(\tilde{X}_{s_a,\,A}^{P_a\phantom \dagger}\tilde{X}_{s_a,\,A}^{P_a\,\dagger}\right)^{-1/2} \otimes\; \tilde{X}_{s_b,\,B}^{P_b\,\dagger} \left(\tilde{X}_{s_b,\,B}^{P_b\phantom \dagger}\tilde{X}_{s_b,\,B}^{P_b\,\dagger}\right)^{-1/2}.
	\label{eq::tensor-product structure of transformation on A and B}
\end{equation}
Then with 
\begin{equation}
	\sum_{i,j}X_{s_a\otimes s_b,i}^\dagger \mathcal{H}_{i,j}X_{j,s_a\otimes s_b} = D_{s_a,\,A} \otimes 1_B + 1_A \otimes D_{s_b,\,B}
	\label{eq::tensor-product structure eigenergies on A and B}
\end{equation}
cluster-additivity of the transformation is a consequence of 
\begin{equation}
\mathcal{A}^\dagger \left(D_{s_a,\,A} \otimes 1_B + 1_A \otimes D_{s_b,\,B}\right)\mathcal{A} = \mathcal{H}_{\mathrm{eff}}^a(A)\otimes 1_B + 1_A \otimes \mathcal{H}_{\mathrm{eff}}^b(B),
\label{eq::cluster-additivity for part a,b of transformation}
\end{equation}
where $\mathcal{A} = \left(\tilde{X}_{s_a\otimes s_b}^{P_n\,\dagger} \left(\tilde{X}_{s_a\otimes s_b}^{P_n\phantom \dagger}\tilde{X}_{s_a\otimes s_b}^{P_n\,\dagger}\right)^{-1/2}\right)$. The transformation as a whole acting on all particle blocks can also be written down and is given as
\begin{equation}
	T^{\mathrm{pca}} = X \left(\sum_m \tilde{X}^{P_m}\right)^\dagger \left(\left(\sum_m \tilde{X}^{P_m}\right)^{\phantom\dagger}\left(\sum_m \tilde{X}^{P_m}\right)^\dagger\right)^{-1/2}.
	\label{eq::projective linked-cluster transformation as a whole acting on full Hilbert space}
\end{equation}
with $\tilde{X}^{P_n} = P_n \tilde{X} P_n$.
\subsection{Explicit form of transformation in terms of projection operators}
It is important to have the transformation also explicitly given in terms of projection operators as this allows for a local expression of the transformation using Kato's formula Eq.\eqref{eq::Kato perturbative expansion of projector on H subspaces} and implies that reduced graph contributions are zero for graphs with more bonds than the perturbation order. For the explicit form we first define
\begin{equation}
	\bar{\mathfrak{R}}_n\equiv\left(\sum_m R_m\bar{R}_mR_m\right)^{-1}\bar{R}_n
	\label{eq::mathfrak R_n operator}
\end{equation}
with 
\begin{equation}
	R_n\equiv \sum_{m<n} P_m.
	\label{eq::definition R_n}
\end{equation}
The transformation then takes the form 
\begin{equation}
	T^{\mathrm{pca}} = \left(\sum_m \left(\bar{P}_m - \bar{P}_m \bar{\mathfrak{R}}_m \right) P_m \right)\left(\sum_m P_m\left(\left(\bar{P}_m - \bar{P}_m \bar{\mathfrak{R}}_m\right)^\dagger \left(\bar{P}_m - \bar{P}_m \bar{\mathfrak{R}}_m\right)\right)P_m\right)^{-1/2}.
	\label{eq::cluster-additive transformation in terms of projection operators}
\end{equation}
To proof the equivalence of \eqref{eq::projective linked-cluster transformation as a whole acting on full Hilbert space} and \eqref{eq::cluster-additive transformation in terms of projection operators} we need to find a way to express $XP_n(\tilde{X}^\dagger-X^\dagger)$ in terms of projection operators. We first note that the conditions $$P_n(\tilde{X}^\dagger-X^\dagger)R_n = - P_n X^\dagger R_n$$ (subtractions of lower-energy states yield $R_n \tilde{X}^{P_n}=0$) and $$P_n(\tilde{X}^\dagger-X^\dagger)\bar{R}_n = P_n\left(\tilde{X}^\dagger-X^\dagger\right)$$ (only states with lower energy than in block $n$ are subtracted) determine $P_n(\tilde{X}^\dagger-X^\dagger)$ uniquely. We need to show that both these conditions are also fulfilled for $-P_n X^\dagger \bar{\mathfrak{R}}_n$ to show that $-\bar{P}_n\bar{\mathfrak{R}}_n = XP_n\left(\tilde{X}^\dagger - X^\dagger \right)$. The latter condition is obviously fulfilled by the construction of Eq.~\eqref{eq::mathfrak R_n operator}. For the first condition we note that 
\begin{equation}
	P_n X^\dagger \bar{\mathfrak{R}}_n R_n= P_n X^\dagger \left(\sum_m R_m\bar{R}_mR_m\right)^{-1}R_n\bar{R}_nR_n = P_nX^\dagger R_n.
	\label{eq::correction of transformation as part }
\end{equation}
This proves the equivalence of Eq.~\eqref{eq::projective linked-cluster transformation as a whole acting on full Hilbert space} and Eq.~\eqref{eq::cluster-additive transformation in terms of projection operators} and establishes the form of the transformation in terms of projection operators only. It is important to have shown this equivalence since perturbatively it follows that one can expand the transformation in local terms using Kato's formula. 
\section{Low-field expansion for transverse-field Ising model on square lattice}
\label{section::Application: low-field expansion for transverse-field Ising model on square lattice}
As an application we investigate the ferromagnetic transverse-field Ising model on the square lattice in the lwo-field ordered phase. The Hamiltonian of this paradigmatic model can be written down with Pauli matrices and takes the form
\begin{equation}
	\mathcal{H} = - \frac{1}{4}\sum_{\langle \nu,\nu^\prime\rangle} \sigma_{z,\nu^{\phantom\prime}} \sigma_{z,\nu^\prime} + h \sum_\nu \sigma_{x,\nu} = \mathcal{H}_0 + h V,
	\label{eq::Hamiltonian transverse-field Ising model}
\end{equation}
with
\begin{equation}
	\mathcal{H}_0 = - \frac{1}{4}\sum_{\langle \nu,\nu^\prime\rangle} \sigma_{z,\nu^{\phantom\prime}} \sigma_{z,\nu^\prime}
	\label{eq::H0 transverse-field Ising model}
\end{equation}
and
\begin{equation}
	V = \sum_\nu \sigma_{x,\nu}.
	\label{eq::perturbation transverse-field Ising model}
\end{equation}
The Hamiltonian commutes with the spin-flip transformation $\prod_{\nu} \sigma_{x,\nu}$. In the ordered phase this $\mathbb{Z}_2$ symmetry is broken and the model undergoes a second-order phase transition in the $3d$ Ising universality class towards the disordered high-field phase when $h$ is increased. Good estimates of the critical point were obtained using high-field series expansions and quantum Monte Carlo simulations and yielded $h_c \approx 0.7610$ \cite{He1990,Hesselmann2016}. Best estimates of the critical exponent can be obtained using the conformal bootstrap method and quantum Monte Carlo simulations \cite{Hasenbusch2010,El-Showk2014}. The first two digits of the correlation length exponent are given as $\nu=0.63$. On finite systems the parity symmetry is not broken. In order to perform linked-cluster expansions one therefore goes into a dual picture that is isospectral to the original one in the infinite system but has a unique polarized ground state for $h=0$. As in \cite{Dusuel2010} we define new pseudo-spin-$1/2$ degrees of freedom and new Pauli matrices 
\begin{equation}
	\tilde{\sigma}_{z,\beta}=\tilde{\sigma}_{z,\langle \nu,\nu^\prime\rangle } = \sigma_{z,\nu^{\phantom\prime}}\sigma_{z,\nu^\prime}
	\label{eq::new spin-z variable dual picture}
\end{equation}
that takes the eigenvalues $\pm 1$ of the Ising interaction on every bond $\langle \nu,\nu^\prime\rangle$. This means that the degrees of freedom are located on the bonds and not on the sites any more. The dual Hamiltonian in this basis can be decomposed into an unperturbed and perturbed part in the following way:
\begin{equation}
	\tilde{\mathcal{H}} = \tilde{\mathcal{H}}_0 + h \tilde{V}
	\label{eq::dual Hamiltonian transverse-field Ising model}
\end{equation}
with 
\begin{equation}
	\tilde{\mathcal{H}}_0 = -\frac{1}{4}\sum_\beta \tilde{\sigma}_{z,\beta}
	\label{eq::H0 of dual Hamiltonian}
\end{equation}
and 
\begin{equation}
	\tilde{V} = \sum_s \tilde{A}_s,
	\label{eq::V of dual Hamiltonian}
\end{equation}
where the plaquette operator $\tilde{A}$ takes the form
\begin{equation}
	\tilde{A}_s=\prod_{\beta\in s(\nu)} \tilde{\sigma}_{x,\beta}.
	\label{eq::plaquette operator dual picture}
\end{equation}
The index $\beta$ runs over the four bonds $s(\nu)$ that are connected to the site $\nu$ in the original degrees of freedom.\\
In this section we are going to employ our transformation $T^{\rm pca}$ to the low-field phase of the model and derive series and NLCE results for the spin-flip and bound-state excitation gap in this model. Bound states arise in this model because flipping two adjacent spins in the ground state yields a state with lower energy in $\mathcal{H}_0$ than flipping two spins further apart. We analyse the series results in the next subsection \ref{subsection::Perturbative results for single spin flip and bound states} and further calculate the same quantities non-perturbatively in subsection \ref{subsection::Non-perturbative results for single spin flip and bound states}. 

\subsection{Perturbative results for single spin flip and bound states}
\label{subsection::Perturbative results for single spin flip and bound states}
Perturbative low-field expansion were most efficiently performed with a transformation of same complexity as the minimal transformation \cite{Oitmaa1991}. Even though this calculation was done on a large number of also non-linked graphs - since it did not allow for a linked-cluster expansion of excitations because of couplings between ground state and excitations - it reached much higher orders than a calculation on only linked-clusters with the pCUT method \cite{Dusuel2010}. Our approach is thus ideal having same complexity as the minimal transformation but allowing for a linked-cluster expansion.\\
We calculated graph embeddings on the square lattice using a hypergraph expansion \cite{Muehlhauser2022} and obtained the embedding factors for all graphs with up to $13$ sites in the original lattice. The elementary excitation in the low-field phase is a spin-flip. Next higher excitations are bound states adiabatically connected to two spin flips on adjacent spins. We calculated the spin-flip gap up to order $24$ extending the results of \cite{Oitmaa1991} by $4$ orders and the bound-state gap up to order $22$ extending the results of \cite{Dusuel2010} by $10$ orders. It is possible to reach such high orders with graphs of only up to $13$ sites since in the low-field expansion of excitations with $a$ spin-flips on a graph with $N$ sites the minimal order for a reduced graph contribution is $2(N-a)$. This property is also called strong-double-touch. We checked that both series agree with the known results of \cite{Oitmaa1991,Dusuel2010}.\\ 
As for our method it is only important to obtain the eigenspaces and energies of the excitation of interest and those of all excitations with lower energy, we used one of the most efficient methods for calculating eigenspaces and energies perturbatively, which is the two-block orthogonalization method (TBOT) form of the minimal transformation. A description of TBOT is given in \cite{Zheng2006}. With the information obtained this way we then construct the cluster-additive projective transformation to perform the linked-cluster expansion for both the spin-flip and bound-state gap. Almost all resources are needed for the TBOT calculation. Hence, we are as efficient as TBOT but only need to consider linked-clusters making the method very efficient.\\
We denote the series for the zero momentum single spin-flip gap by $\Delta$ and the one for the zero momentum bound-state gap by $\Delta_{\mathrm{bs}}$. They read respectively
\begin{equation}
	\begin{aligned}
	\Delta & = 2 - 3\,h^2 + 3.5833\,h^4 - 23.140\,h^6 + 133.22\,h^8 - 849.05\,h^{10} + 5738.0\,h^{12} \\
	& - 40573\,h^{14} + 29615\cdot 10\, h^{16} - 22157\cdot 10^2\, h^{18} + 16906\cdot 10^3\, h^{20}\\
	& - 13105\cdot 10^4\, h^{22} + 10292\cdot 10^5\,h^{24}
	\end{aligned}
	\label{eq::series Delta 2}
\end{equation}
and 
\begin{equation}
	\begin{aligned}
		\Delta_{\mathrm{bs}} & = 3 - 22.916\,h^4 - 13.334\,h^6 + 263.64\,h^8 + 5213.1\,h^{10} - 7214.0\,h^{12} - 31023\cdot 10 \,h^{14} \\
		& - 24296\cdot 10^2 \,h^{16} + 19814\cdot 10^3 \, h^{16} + 30204\cdot 10^4\, h^{20} + 57170\cdot 10^4\, h^{22}.
	\end{aligned}
	\label{eq::series Delta 3}
\end{equation}
Note that we displayed the first five digits of the coefficients and did not round to the last digit. This accuracy can be guaranteed, while for more digits calculations would have needed to be performed with higher accuracy than double precision.\\
 To analyse the behaviour of these series we used Pad\'{e} and DLog-Pad\'{e} extrapolations. A good and extensive review about extrapolation techniques in general and especially these two is \cite{Guttmann1974}. Pad\'{e} approximations are a well established tool to enhance the convergence of a perturbative series and DLog-Pad\'{e} extrapolations in particular mimic the algebraic behaviour of critical quantities in the vicinity of a quantum phase transition.\\
The series $\Delta$ of the gap is consistently alternating up to high orders. Many DLog-Pad\'{e} extrapolations of $\Delta$ break down because of spurious poles. To estimate the reliability of DLog-Pad\'{e} extrapolations it is helpful to study the convergence behaviour of the  DLog-Pad\'{e} families of order $[n,n+d]$ with $d$ fixed. We found that only the families with $d=\pm 2$ show converging behaviour and that the family $d=2$ looks better converged. The critical point of the extrapolation of the highest order, i.e. the $[10,12]$ DLog-Pad\'{e} extrapolant, yields a critical point of $h_c=0.762$ and a critical exponent of $\nu=0.649$.\\ 
An extrapolation analysis of $\Delta_{\mathrm{bs}}$ is in principle also reasonable as the bound-state mode is stable and expected to close with the same critical exponent as the spin-flip gap, i.e. $\nu(\Delta) = \nu(\Delta_{\mathrm{bs}})$. Indeed, there are field theoretic calculations of Caselle et al. \cite{Caselle2000,Caselle2002} predicting $\Delta_{\mathrm{bs}}/\Delta \vert_{h=h_c} \approx 1.8 $. This quantity was also calculated with exact diagonalisation yielding a value of $1.84(3)$ \cite{Nishiyama2008}. Unfortunately, the series of the bound state $\Delta_{\mathrm{bs}}$ shows a complicated behaviour and no convergence of Pad\'{e} or DLog-Pad\'{e} extrapolations was found. In \cite{Dusuel2010} $\Delta_{\mathrm{bs}}/\Delta$ was investigated with Pad\'{e} and DLog-Pad\'{e} extrapolations but only one extrapolation, the DLog-Pad\'{e} $[4,6]$, showed non-spurious behaviour and a value close to the numerical value of  $1.84(3)$ as in \cite{Nishiyama2008}. Having calculated ten orders of perturbation more than in \cite{Dusuel2010} one could hope that we find more extrapolations consistent with the predictions and calculations of \cite{Caselle2000,Caselle2002,Nishiyama2008}. However, this is not the case and the additional orders rather show that the DLog-Pad\'{e} family of the DLog-Pad\'{e} $[4,6]$ extrapolant does not seem to converge with higher orders. At least up to the calculated orders so far, no behaviour of the series extrapolations that is consistent with the expectation of $\Delta_{\mathrm{bs}}/\Delta \vert_{h=h_c} \approx 1.8 $ could be found.

\subsection{Non-perturbative results for single spin flip and bound states}
\label{subsection::Non-perturbative results for single spin flip and bound states}
\begin{figure}[!htbp]
	\centering	
	\includegraphics[width=0.8\textwidth]{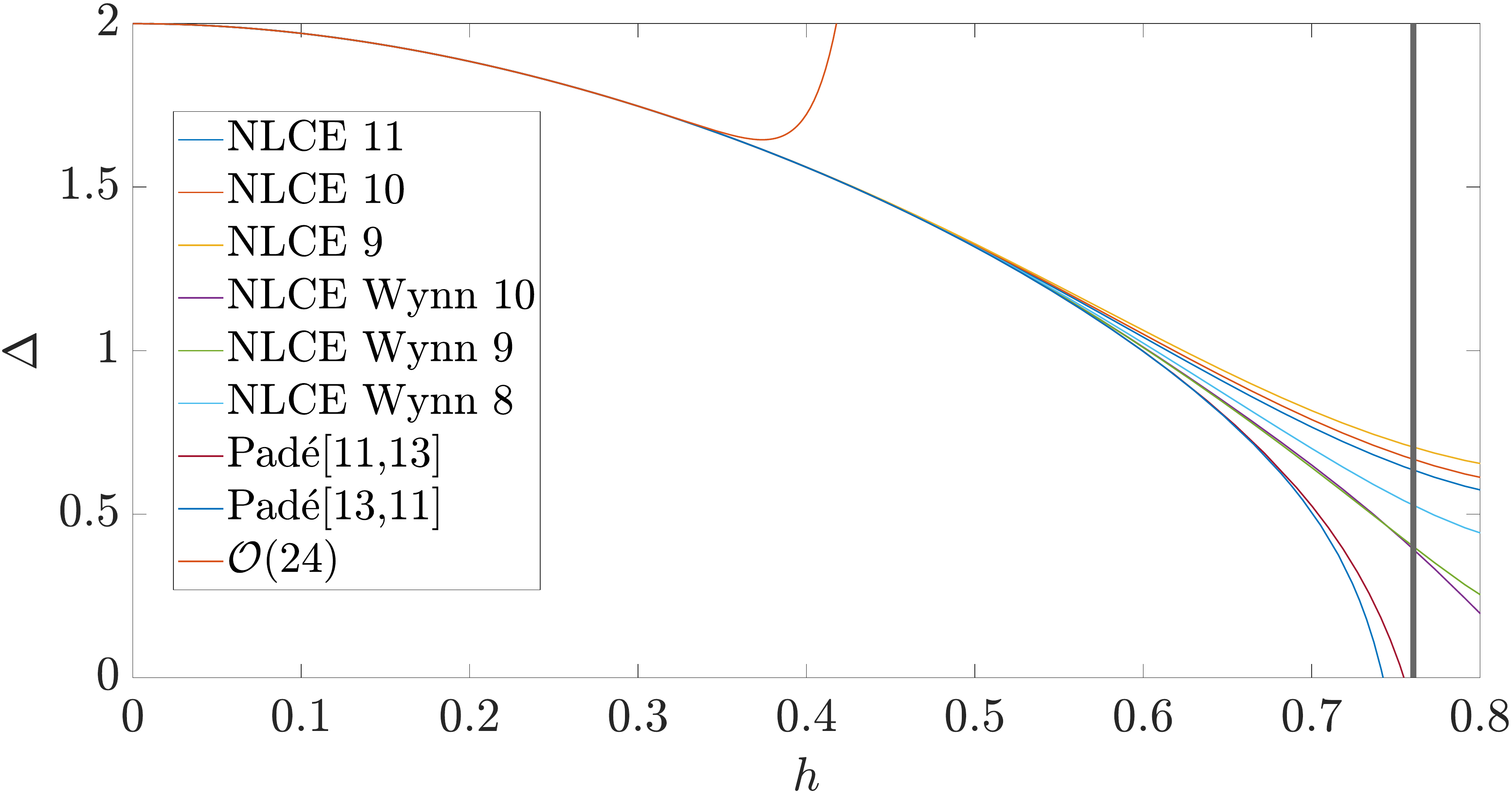}
	\caption[NLCE expansion spin-flip gap $\Delta$]{The figure shows an NLCE expansion of the spin-flip gap $\Delta$ in dependence of the number of vertices of the graphs taken into account. The expansion converges until around $h\approx 0.5$. The phase transition point $h_c \approx 0.7610$ \cite{He1990,Hesselmann2016} is highlighted as a black vertical line. Wynn extrapolations of the NLCE expansion converge up to slightly larger values of $h$ but converge only slowly towards the critical point. Pad\'{e} extrapolations are also shown together with the bare series.}
	\label{fig::NLCE_SpinFlip_11Nodes}
\end{figure}

\begin{figure}[!htbp]
	\centering	
	\includegraphics[width=0.8\textwidth]{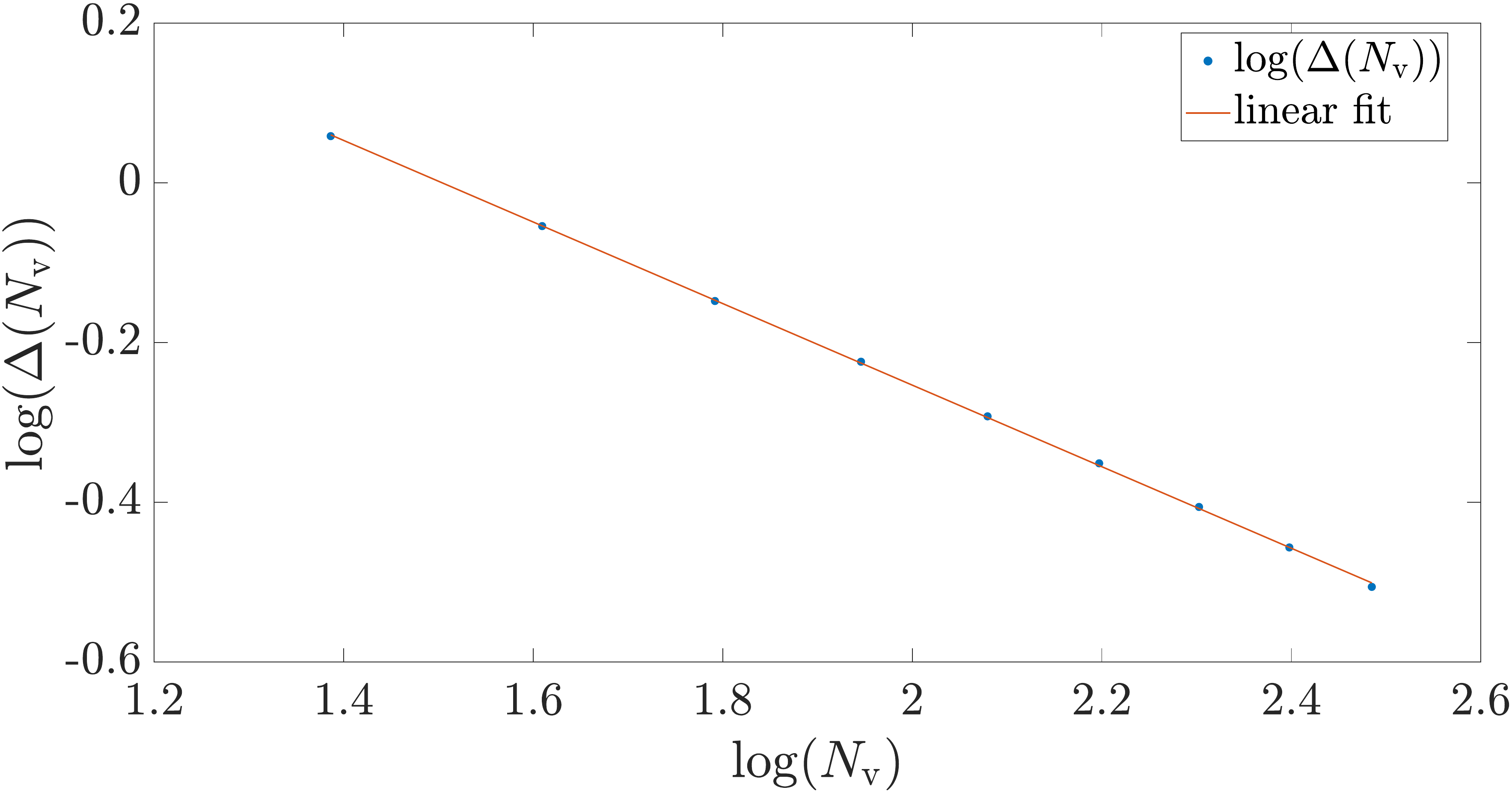}
	\caption[Scaling of $\Delta(N_{\mathrm{v}})$ at the critical point $h_c$]{The plot shows the scaling of the energy gap $\Delta$ in the dependence of the maximum number of vertices $N_{\mathrm{v}}$ of graphs used in the NLCE in a double-logarithmic plot. A linear fit of good quality shows that the behaviour is algebraic with an exponent of $\kappa=-0.51$.}
	\label{fig::NLCE gap fit at critical point}
\end{figure}

\begin{figure}[!htbp]
	\centering	
	\includegraphics[width=0.8\textwidth]{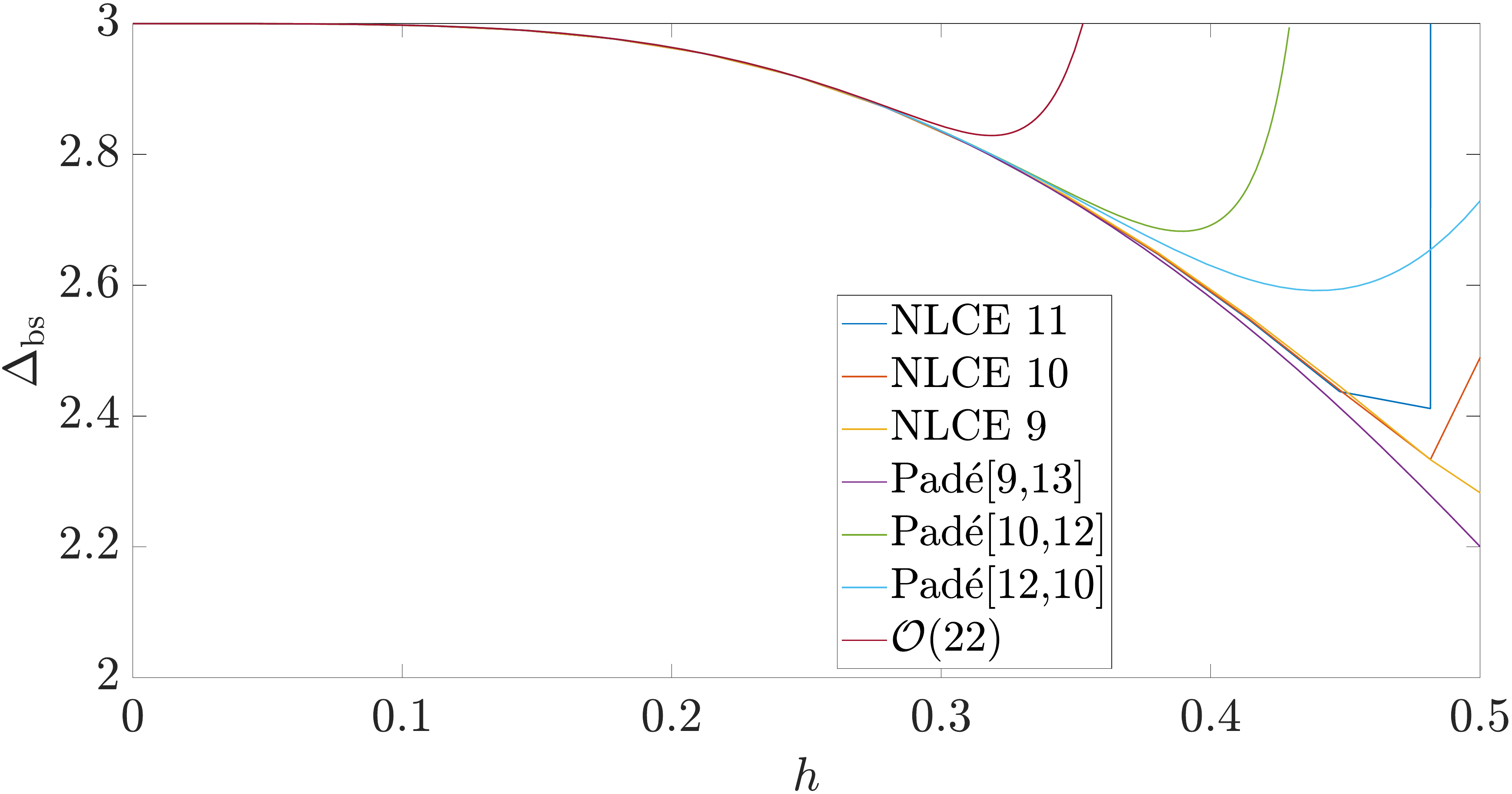}
	\caption[NLCE expansion bound-state gap $\Delta_{\mathrm{bs}}$]{The figure shows an NLCE expansion of the bound-state gap $\Delta_{\mathrm{bs}}$ in dependence of the number of nodes of the graphs taken into account. The expansion converges only until around $h\approx 0.35$. The convergence problems are caused by avoided level crossings occurring on finite graphs. As more graphs are taken into account in the expansion convergence becomes gradually worse. Pad\'{e} extrapolations and bare series results are also shown.}
	\label{fig::NLCE_BoundState_11Nodes}
\end{figure}

\begin{figure}[!htbp]
	\centering	
	\includegraphics[width=0.8\textwidth]{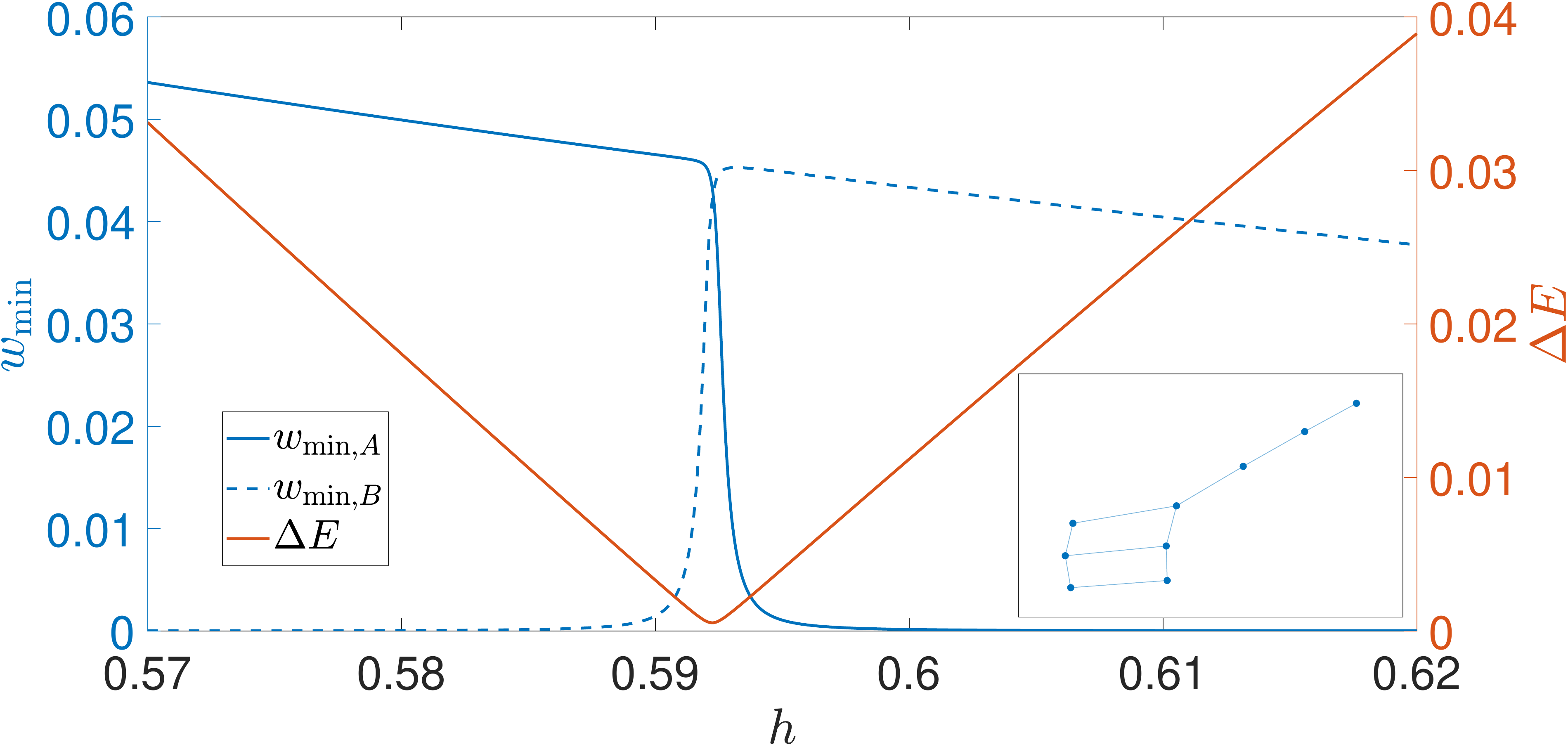}
	\caption[Signatures of avoided level crossing]{The figure shows the behaviour of the minimal eigenvalue $w_{\mathrm{min},A}$ of $W_2$ (blue line) in the vicinity of an avoided level crossing for the calculation of the effective Hamiltonian on a finite graph, which is plotted in the inset of the figure. In the same plot the energy difference $\Delta E$ between the lower end of the two-spin flip continuum and the maximum of the bound-state dispersion is plotted (red). One clearly recognizes that $w_{\mathrm{min},A}$ drops to a very small value as $\Delta E$ decreases. As a blue dashed line the minimal eigenvalue $w_{\mathrm{min},B}$ of a modified $W_2$ is shown, where one takes the formerly lower two-spin flip continuum continuum state for the calculation of the bound-state effective Hamiltonian and rejects the state that was formerly the one with highest energy of the bound states. The plot clearly suggests further away from the avoided level crossing the dashed blue curve would continue the solid blue one smoothly.}
	\label{fig::Min_EW_Alc}
\end{figure}

Non-perturbative linked-cluster expansions (NLCEs) for the low-field phase of the transverse-field Ising model were so far only performed for ground-state energies and ground-state expectation values of observables \cite{Ixert2016,Thompson2018}. In these papers the linked-cluster expansion for the ground state was not performed in the dual picture but in a more optimised setting to capture fluctuations of the environment that act back onto the closed finite system of a graph. Here we stay in the dual picture because a modified coupling due to the environment is not obvious for excited states. With NLCEs one can obtain converging results for larger values of $h$ than with perturbation theory. As long as the correlation lengths are captured within the length scale of graphs considered it is reasonable to assume that NLCEs can converge. In contrast to perturbative expansions where order of perturbation and length scales are coupled, for NLCEs this is not the case any more since an exact calculation on a graph can be thought of as a resummation of an infinite order expansion on that graph. Consequently, convergence properties of both approaches can be different.\\
With the NLCE applying our transformation $T^{\rm pca}$ we also calculated $\Delta$ and $\Delta_{\mathrm{bs}}$ using exact diagonalisations with ARPACK routines to obtain the low-energy spectrum and eigenvectors of $\mathcal{H}$. In Fig.~\ref{fig::NLCE_SpinFlip_11Nodes} we show plots of the spin-flip gap for different numbers of vertices of the graphs used in the expansion and compare with extrapolations of the series results. The NLCE converges to values of $h\approx 0.5$ extending the convergence of the bare series. We also show Wynn extrapolations \cite{Brezinski2000} with regards to the number of nodes of graphs in Fig.~\ref{fig::NLCE_SpinFlip_11Nodes}. Wynn extrapolations of a series $S_o$ depending on an expansion parameter $o$ are defined as 
\begin{equation}
	\frac{S_{o+1}S_{o-1}-S_o^2}{S_{o+1}-2S_0+S_{o-1}}.
	\label{eq::Wynn extrapolations}
\end{equation}
These extrapolations extend the convergence of the NLCE a bit further but it still breaks down before the critical point at $h_c \approx 0.7610$ \cite{He1990,Hesselmann2016}. One way to access critical exponents with NLCEs is to scale the spin-flip energy gap with respect to the number of vertices $N_v$ of graphs used in the expansion at the position $h_c \approx 0.7610$ of the estimated critical point. A logarithmic plot of this is shown in Fig.~\ref{fig::NLCE gap fit at critical point} together with a linear fit. This fit yielded an exponent of $\kappa = -0.51$. As in this model one would expect the gap to scale with the inverse correlation length this result implies that not the number of vertices $N_{\mathrm{v}}$ but the square root of it scales in the same way as the correlation length. Although this analysis does not allow for a very precise determination of the critical point it clearly is consistent with a critical value of $h_c \approx 0.7610$ and hence shows that critical behaviour can be captured with NLCEs of excitation gaps.
\\
The NLCE expansion of the bound-state gap converges up to $h \approx 0.35$. For a perturbative calculation of the bound-state energy it does not matter if one subtracts only the ground-state parts from the bound-state eigenvectors or both the ground-state and single-spin flip part as described in Eq.~\eqref{eq::transformed tilde psi states}. Interestingly, the NLCE broke down earlier when only the ground-state part was subtracted so we always also subtracted the spin-flip part. Results are shown in Fig.~\ref{fig::NLCE_BoundState_11Nodes}. The reason for worse convergence in comparison to $\Delta$ is energetic overlap between bound states and the two-spin flip continuum \cite{Dusuel2010}. This is a well known problem in all sorts of effective Hamiltonian theories and for example also shows up in quantum chemistry as intruder state problem on finite systems \cite{Malrieu1985} or in graph-based continuous unitary transformations (gCUT) \cite{Coester2015a}. Only a finite number of eigenstates and eigenvectors exists in a finite system. Energetic overlap between two different sorts of formerly gapped quasi-particles shows up as an avoided level crossing. These avoided level crossings are also connected to exceptional points in the complex plane of the perturbation parameter that we follow adiabatically \cite{Heiss1990}. As pointed out in \cite{Malrieu1985} either one follows adiabatically the low-lying state and looses transferability of the expansion or one tracks the right states but then has a problem of smoothness of the expansion around the avoided level crossing. A promising solution to overcome this problem was found in \cite{Coester2015a}, where in the region of an avoided level crossing not exact but only approximate eigenstates where used to track the right diabatic states as good as possible and not the adiabatic ones any more. They used continuous unitary transformations based on the quasi-particle generator in Eq.~\eqref{eq::QP generator} \cite{Knetter2000} but using a modified generator around the anti-level crossing. Next to observable characteristics they took a quantity known from the CORE method as characteristic to identify such pseudo-particle decay. For single-particle excitations not coupled to the ground state this quantity behaves similar as the minimal eigenvalue of Eq.~\eqref{eq::particle decay characteristic generalized}
\begin{equation*}
	W_n=\sum_{m<n+1}P_m \sum_{m<n+1}\bar{P}_m  \sum_{m<n+1}P_m.
	\label{eq::particle decay characteristic generalized}
\end{equation*}
While a generalization to the generic case seems not so clear within the CORE approach $W_n$ naturally shows up in our approach and can be used to identify particle-decay of higher energetic excitations or excitations coupled to the ground state. Indeed, Fig.~\ref{fig::Min_EW_Alc} shows a graph where avoided level crossings related to the quasi-particle decay occur. As can be seen, the minimal eigenvalue $w_{\mathrm{min}}$ of Eq.~\eqref{eq::particle decay characteristic generalized} drops to zero as the two eigenvalues of the bound states and spin-flip states approach each other. While decay is expected for high-energy momentum modes in the thermodynamic limit the low-energy modes of the bound states are expected to remain stable. Hence, it could be possible to keep some decay channels open but to still do a linked-cluster expansion for the stable bound-state modes. A solution to this problem in our approach could be to not use exact projective eigenspaces around an avoided level crossing but only approximate eigenspaces in the spirit of \cite{Coester2015a}, still demanding pairwise orthogonality of each space. A treatment of this problem is beyond the scope of this paper. We stress that it is not clear if a parameter-free or even cluster-additive solution to this problem exists in general.

\section{Conclusions}
\label{section::Conclusions}
We described how to construct a cluster-additive transformation for excitations of a Hamiltonian $\mathcal{H}=\mathcal{H}_0+\lambda V$ with energies $e^n$ adiabatically connected to the energies $e_0^n$ of $\mathcal{H}_0$. The transformation only depends on the projectors of eigenspaces $e_0^m\leq e_0^n$ of $\mathcal{H}_0$ and the projectors of the adiabatically connected eigenspaces of $\mathcal{H}$. In that respect the transformation needs minimal information content compared to other genuine cluster-additive transformations while generalizing the well known minimal transformation, which uses projectors on the eigenspace $e_0^n$ and the adiabatically connected space of $\mathcal{H}$ only, but is not cluster-additive in general. We also give the transformation explicitly in terms of projection operators, which implies basis independence and local expressibility of the perturbative expansion following from the projector expansion of Kato \eqref{eq::Kato perturbative expansion of projector on H subspaces}. As an application we performed a low-field linked-cluster expansion for spin-flip and two spin-flip bound state excitations in the transverse-field Ising model on the square lattice. We did this both perturbatively and non-perturbatively. \\
Both in the perturbative and non-perturbative setting the method is computationally very efficient. The complexity for perturbative calculations is similar to the TBOT method, which is the most efficient method for high-order matrix perturbation theory we know of. Non-perturbatively the complexity is that of Krylov-based diagonalisation methods. While perturbatively it is hard to come up with further improvements of the method, in non-perturbative applications using exact eigenvectors of finite-lattice Hamiltonians problems arising in the vicinity of avoided level crossings still present a major obstacle. Promising approaches to overcome this problem were given in \cite{Coester2015a}. To find a parameter-free and cluster-additive way of dealing with avoided-level crossings in the construction of effective Hamiltonians remains an important task for the future. If this is achieved the proposed transformation provides a highly efficient tool to perform linked-cluster expansions for excitations in generic Hamiltonians with the possibility to describe decay of excitations accurately and efficiently.\\
We want to end the paper with possible applications of the introduced method. The minimal transformation only allows for a perturbative linked-cluster expansion of excitations that are in a different symmetry sector than the ground state. In almost all low-field expansions this is not the case. While it is possible to perform such expansions with pCUT or MBOT these methods are less efficient than the method we propose. Hence, it promises to reach higher orders in low-field expansions in general, what we already showed specifically for the transverse-field Ising model on the square lattice. High-field expansions of models where the ground state is coupled with the first excited states can also be computationally very demanding. An example is the Kitaev model in a field \cite{Jahromi2021,Schellenberger2022}. The proposed transformation could help to reach higher orders for that system. Another advantage compared to pCUT is that we do not need an equidistant spectrum of $\mathcal{H}_0$. In \cite{Coester2015} it was proposed to use the model independent structure of the pCUT solution to treat systems with disorder or long-range interacting systems and this idea, coined white-graph expansion, was also successfully applied \cite{Fey2016,Hormann2018}. Using perturbative expansions of projectors we can do the same with this transformation but in a more general setting of non-equidistant $\mathcal{H}_0$. This can be utilized to perform white-graph expansions for the resolvent revealing the possibility of long-range low-field linked-cluster expansions and low-field linked-cluster expansions in the presence of quenched disorder.


\section*{Acknowledgements}

MH thanks Matthias M\"uhlhauser for fruitful discussions as well as for embeddings and graphs for the low-field expansion of the transverse-field Ising chain. This work was funded by the Deutsche Forschungsgemeinschaft (DFG, German Research Foundation) - Project-ID 429529648 - TRR 306 QuCoLiMa (Quantum Cooperativity of Light and Matter).
KPS and MH acknowledges the support by the Munich Quantum Valley, which is supported by the Bavarian state government with funds from the Hightech Agenda Bayern Plus.

\end{document}